\documentclass[twocolumn,showpacs,aps]{revtex4}
\usepackage{amsmath,amsthm,amscd}
\usepackage{dcolumn}
\usepackage{graphicx}
\usepackage{epsfig}
\begin{document}
\newcommand{\Qb}{\ensuremath{\mathbf{Q}}}
\newcommand{\qb}{\ensuremath{\mathbf{q}}}
\newcommand{\Sb}{\ensuremath{\mathbf{S}}}
\title{Comment on `Distorted perovskite with $e_g^1$ configuration as a frustrated spin system'}
\author{T. A. Kaplan and S. D. Mahanti}
\affiliation{Department of Physics \& Astronomy and Institute for Quantum Sciences, Michigan State University\\
East Lansing, MI 48824}
\begin{abstract}
Interesting magnetic structure found experimentally in a series of manganites RMnO$_3$, where R is a rare earth,
was explained, ostensibly, by Kimura et al, in terms of a classical Heisenberg model with an assumed set of
exchange parameters, nearest-neighbor and next-nearest-neighbor interactions, $J_n$. They calculated a phase
diagram as a function of these parameters for the ground state, in which the important ``up-up-down-down" or
E-phase occurs for finite ratios of the $J_n$. In this Comment we show that this state does not occur for such
ratios; the error is traced to an incorrect method for minimizing the energy. The correct phase diagram is
given. We also point out that the finite temperature phase diagram presented there is incorrect.
\end{abstract}
\maketitle

Interesting magnetic phases that had been observed in a series of manganites RMnO$_3$, with R=holmium in
particular, were explained in the cited paper~\cite{kimura} on the basis of a Heisenberg Hamiltonian with
classical spins on a 2-d rectangular lattice with ferromagnetic nearest-neighbor exchange parameter,  $J_1$, and
two different antiferromagnetic next-nearest- neighbor parameters, $J_2$ and $J_3$. (The difference is due to
the asymmetric positions of the oxygens resulting from Jahn-Teller distortion). By considering a small set of
states including the unusual E-type, or ``up-up-down-down" state, they obtained a phase diagram for the ground
state, and at finite temperature T. We show that the ground state phase diagram is incorrect, and point out that
the finite-T case is also incorrect, the correct results differing qualitatively from those presented there. The
importance of these conclusions is heightened by the fact that detailed results of the model calculation are
accepted in~\cite{dong}.

Circa 1960 it was shown that guessing various spin states (``reasonable", or experimentally determined) as
ground states of the classical Heisenberg energy,
\begin{equation}
H=\sum J_{ij}\Sb_i\cdot\Sb_j, \mbox{where}\  \Sb_i^2=1,\label{1}
\end{equation}
with some assumed set of exchange parameters $J_{ij}$, and comparing their energies to arrive at a phase
diagram, is fraught with danger, often giving incorrect results. (The spin lengths have been absorbed into the
$J's$.) Furthermore, for spins on a Bravais lattice, it was shown~\cite{lyons1} that the ground state energy is
always attained by a simple spiral with the wave vector $\Qb$ that minimizes $J(\qb)$, the Fourier transform of
of the exchange parameters $J_{ij}$ (the LK theorem). The energy per spin of a spiral with wave vector $\qb$ is
just $J(\qb)$. The spiral concept was introduced in 1959.~\cite{yoshimori, kaplan1, villain} This is relevant to
the case discussed in~\cite{kimura}, the lattice considered being a Bravais lattice (one spin per primitive unit
cell). The proof of the theorem was via the Luttinger-Tisza method~\cite{luttinger1,luttinger2,luttinger3};
please see the recent review~\cite{kaplan2}.
\begin{figure}[h]
\includegraphics[height=8in]{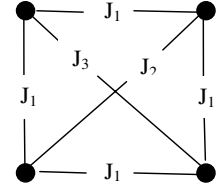}\vspace{-420pt}
 \caption{Unit cell showing the exchange interactions between the Mn ions.}
 \label{fig:unit cell}
\end{figure}

The case under consideration~\cite{kimura} essentially involves spins on a square lattice, the 4 n.n.
interactions being equal ($J_1$), but, due to the internal oxygen structure, the n.n.n interactions $J_2, J_3$
are different. The unit cell is shown in Fig. 1, where the $J's$ are indicated (the oxygens are not shown).
Taking the x and y axes respectively, pointing to the right and up with respect to the figure, one readily finds
\begin{equation}
J(\qb)/2=J_1(\cos q_x+\cos q_y)+J_2\cos(q_x+q_y)+J_3\cos(q_x-q_y).\label{2}
\end{equation}
In~\cite{kimura}, $J_1=-|J_1|$ (ferromagnetic n.n. interactions), $J_2>0$ and $J_3$ is neglected, so that
\begin{equation}
J(\qb)/(2|J_1|)=-(\cos q_x+\cos q_y)+\gamma\cos(q_x+q_y),\label{3}
\end{equation}
where $\gamma=J_2/|J_1|$. We first consider $\qb$ in the (1,1) direction, $\qb=p(\hat{x}+\hat{y})$, so, putting
$|J_1|=1$,
\begin{equation}
J(\qb)/2=-2\cos p+\gamma\cos(2p).\label{4}
\end{equation}
The stationary points of this are given by $\sin p=0$ and
\begin{equation}
\cos p=1/(2\gamma)\equiv\cos p_0;\label{5}
\end{equation}
the latter occurs only if $2\gamma\ge 1$ (the wave vector is real.) The energies at $\qb=0$ and at
$p_0(\hat{x}+\hat{y})\equiv\Qb$ are, from~(\ref{4}) and (\ref{5}),
\begin{eqnarray}
J(0)/2&=&-2+\gamma\nonumber\\
J(\Qb)/2&=&-\frac{1}{\gamma}+\gamma(\frac{1}{2\gamma^2}-1)=-\frac{1}{2\gamma}-\gamma.\label{5}
\end{eqnarray}
These are equal at $\gamma=1/2$ and in fact
\begin{equation}
J(0)/2-J(\Qb)/2=2(\gamma+\frac{1}{4\gamma}-1)=\frac{1}{2\gamma}(2\gamma-1)^2;\label{6}
\end{equation}
that is, the spiral (see the LK theorem) goes below the ferromagnetic state as $\gamma$ increases past 1/2. This
contradicts the story in~\cite{kimura} where the ferromagnetic state is indicated to be the ground state all the
way to $\gamma=1$. Furthermore, the ground state is not the `up-up-down-down" configuration when the
ferromagnetic state is first destabilized, as claimed in~\cite{kimura}, but rather a spiral whose wave vector
increases continuously from zero at $\gamma=1/2$. Interestingly, as $\gamma\rightarrow\infty,
p_0\rightarrow\pi/2$, which gives the spiral with a 90$^\circ$ turn angle; and as pointed out in~\cite{dong},
this is degenerate with the ``up-up-down-down" state. These results are summarized in Fig. 2.
\begin{figure}
\includegraphics[height=3in]{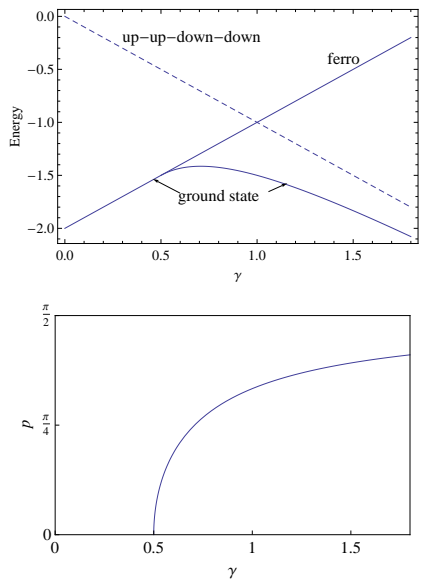}
 \caption{upper: Energy of the ground and other states. lower:  Component $p_0$ of the spiral wave vector.}
 \label{fig:energies}
\end{figure}

To complete the solution to the ground state of this model, we need to check that the lowest $J(\qb)$ that we
found (under the restriction $\qb$ in the (1,1) direction), is lowest over all $\qb$ in the Brillouin zone. We
did this for two values of $\gamma$, namely, 0.4 and 0.6 (sample values in the ferromagnetic region and the
spiral region, respectively). Indeed the (1,1) solution is the minimum over all $\qb$. We consider this
sufficient, although it is easy to check further values of $\gamma$. Lest one might think that we have merely
found the lowest spiral (the ferromagnet is the $\qb=0$ spiral), and hence ask whether there are other
lower-energy states with more than one wave vector (and its negative) in their Fourier representation, we note
that because of the place of the calculation in the framework of the Lutttinger-Tisza method and the LK theorem,
the conclusion is far stronger: \emph{there is no state with a lower energy than the solution
given}.~\cite{konstantinidis} The finite-T phase diagram in~\cite{kimura}, obtained within the mean-field
approximation, shows very complex behavior, numerous long-range-ordered phases (the ``Devil's flower"). However,
it was shown by Lyons~\cite{lyons2}, that for a Bravais lattice with T-independent exchange parameters (the case
considered in~\cite{kimura}), the phase boundary is vertical: a spiral at the ordering temperature remains a
spiral with the same wave vector for all T. That is, all the sweet complexity shown in the phase diagram
of~\cite{kimura} is not a property of their assumed Hamiltonian.

Thus the question remains, what is the source of the E- or ``up-up-down-down" phase found in HoMnO$_3$? It seems
to us that the answer should be found in the standard fundamental model of localized electrons in this
insulator, as set forth by P.W. Anderson~\cite{anderson}, and characterized qualitatively for the Heisenberg
interactions that are a consequence~\cite{anderson} by the Goodenough-Kanamori rules (see\cite{anderson}). That
will include Heisenberg terms with more general interactions than assumed so far, as well as anisotropic terms
and isotropic higher-spin terms, e.g. $(\Sb_1\cdot\Sb_2)^2$, etc. One important aspect of anisotropy is that it
can remove the degeneracy of the 90$^\circ$ spiral and the E-state (this observation renders incorrect the
argument in~\cite{dong} leading to their conclusion that the ``classical spin model is not suitable to explore
in a single framework the many phases found in RMnO$_3$ manganites").

\end{document}